\documentstyle[12pt]{article}
\begin{document}

\title{A comparison of FQHE quasi-electron trial wave functions on the disk }

\author{M. Kasner and W. Apel \\Physikalisch-Technische
Bundesanstalt
\\Bundesallee 100
\\D--38116 Braunschweig, Germany}
\date{}
\maketitle
\begin{abstract} The various proposals for FQHE quasi-electron trial
wave functions are reconsidered. In a short-range
model for the electronic interaction, the energy expectation values of four
different trial wave functions are calculated at filling factor $\frac{1}{3}$
 for up to
ten electrons in the disk geometry. Jain's trial wave function
displays the lowest energy expectation value.
\end{abstract}
PACS-Numbers: 73.20Dx, 73.50Jt
\newpage
It is by now generally accepted that the fractional quantum Hall effect (FQHE)
arises due to peculiar features of the
multiparticle energy spectrum of interacting, spin-polarized  electrons moving
in two dimensions
under the influence of a strong, perpendicular magnetic field \cite{Mac92}.
These peculiarities appear at definite filling factors $\nu$ (electronic
densities).
Best understood are the ground states
at filling factors $\nu =\frac{1}{q}$ ($q$ - odd).
Here, the Laughlin wave function
\cite{Lau83b} is not only a good trial wave function, but also
an exact solution of a special short-range two-particle interaction
\cite{Hald83}. Quasi-particles were introduced in order to describe the
low-lying
energy levels in systems with a few flux quanta off from
$\nu =\frac{1}{q}$.
They were also constructed in order to explain the
occurence of the other filling factors $\nu =\frac{p}{q}$ by the concept of a
condensation of these
quasi-particles, quite analogous to the condensation of the interacting
electrons into the $\frac{1}{q}$-state \cite{Halp84}.
However, this picture is based on some prepositions. These are the dominance
of the short-range contribution in the pair-interaction of the
quasi-particles and the smallness of the
interaction between these quasi-particles
in comparison with the energy gap of the collective excitations of the parent
state \cite{Hald90}. Also a microscopic derivation of the quantum
mechanics of these quasi-particles from first principles is still missing. \\
The theory received renewed interest when a different approach towards an
explanation of filling factors unequal $\frac{1}{q}$ was proposed by Jain
\cite{Jai89b}, who
gave an explicit prescription not only for constructing trial wave functions
for any number of quasi-particles at $\nu =\frac{1}{q}$ but also for
ground states
 and even the lowest excitations at all the other filling factors. The idea was
to exploit the
incompressibility of non-interacting electrons at integer filling
factors. A multiplication with a symmetric Jastrow-factor then leads to a
correlated wave
function which has the desired filling
factor.
Finally, the wave function is projected onto the lowest Landau
level $n=0$, an almost unnecessary procedure, since after the multiplication
with the Jastrow-factor, only a little weight remains in higher Landau levels
\cite{TJ91}. This formulation was successfully used to explain
 multiparticle spectra on the sphere \cite{DJ92b}. On the other hand, there
are many quantitative results
supporting the previous hierarchy scheme \cite{BM91}.
However, the relation between
both approaches is still unclear.\\
In order to study the relative merits of these two approaches, we choose
the
simplest case where differences can appear. While the construction of
the $\nu =\frac{1}{3}$-quasi-hole leads in either case to the same trial wave
function, the
quasi-electron of the $\frac{1}{3}$-state given by Jain differs from the
quasi-electron proposed
originally by Laughlin. It is not surprising that, in contrast to the case of
the quasi-holes,
there are many proposals for different quasi-electron trial wave functions
\cite{MG86a,MH86,KA92}.\\
The Laughlin wave function is quite a good trial wave function
for various two-particle interactions, and this expresses the universality of
the FQHE. Nevertheless, the characteristic features of the ground state can
be destroyed, if
the short-range contribution no longer dominates the interaction
\cite{Hald90}. Therefore, and because the Laughlin wave function is an exact
solution to this, we use the short-range model as a canonical model. \\
In a first attempt, we aim at the properties of the
$\nu =\frac{1}{3}$-quasi-electron.
Here, we work in the disk geometry \cite{KA93} which was
originally used
to formulate Laughlin's idea, although we are aware of some drawbacks
in comparison with
the spherical geometry.\\
Crucial for the quasi-electron
construction in the hierarchical scheme is that all angular momentum
components of the quasi-electron are degenerated in energy.
On the sphere, this is a priori fulfilled
because of the symmetry. For a system in the disk geometry with a finite
particle number $N$, this is not satisfied, but in the thermodynamic limit the
energies for all the components of a trial quasi-electron
wave function should tend to the same limiting value. Deviations from the limit
can be assigned to i) the quality of the proposed trial wave function and,
of course, ii) to
the influence of the finite size of the system. The finite size
corrections of the ground state energy, e.\ g.\,
are proportional to $\frac{1}{\sqrt{N}}$ on a disk and proportional to
$\frac{1}{N}$ on a
sphere.  It is our aim to study the full dependence of the quasi-electron
energies on the angular momentum for finite $N$.
In previous studies of the disk geometry,
the quasi-electron at the origin was considered corresponding to just a
single angular momentum component. \hspace{1.0cm}   \\
The Hamiltonian to be investigated describes $N$
spin-polarized electrons moving
on a two-dimensional
disk. The Hilbert space is restricted to the lowest Landau
level and thus contains just an interaction term because we do not include a
background in our model. It reads:
\begin{equation}
H=\frac{1}{2} \sum_{m_{1},m_{2},m_{3},m_{4}=0}^{m_{max}}
W_{m_{1}m_{2}m_{3}m_{4}}
c_{m_{1}}^{+}c_{m_{2}}^{+} c_{m_{3}} c_{m_{4}},
\end{equation}
where $c_{m}^{+}$ creates an electron
state with angular momentum $m$ in the lowest
Landau level ($m=0,1,\ldots$). The matrix elements $W_{m_{1}m_{2}m_{3}m_{4}}$
are specified such that only the pseudo-potential coefficient
$V_{1}=\frac{\sqrt{\pi }}{4}$ is nonzero \cite{Hald90}, i.\ e.\ only
electrons with relative angular momentum one repel each other. \\
The calculations are performed for the quasi-electrons of the
$\frac{1}{3}$-state.
Because the filling factor is defined by the relation
$\nu =\frac{N-1}{m_{max}}=\frac{N-1}{N_{\Phi}-1}$, where $m_{max}$ is the
maximum single particle angular momentum and limits the area of the disk
and where
$N_{\Phi }$ is the number of flux-quanta through this disk, the
relation $m_{max}=3(N-1)$ holds for the stable state at $\nu =\frac{1}{3}$.
If we increase the density of the system by keeping $N$ fixed while
decreasing the area by the area of just one flux quantum, we create a
quasi-electron.
Thus, for a
one-quasi-electron trial wave function at $\nu =\frac{1}{3}$ we must have
\begin{equation}
m_{max}=3(N-1)-1.
\end{equation} \\
In this work we report calculations of the energy
expectation values of the angular momentum components for four different
quasi-electron
wave functions proposed in the literature. This is done for up to
$N=10$ particles. \\
The quasi-electron wave functions have total angular momenta $M$ reaching
from $M^{*}-N$ to $M^{*}-1$, where $M^{*}=\frac{q}{2}N(N-1)$
is the total angular momentum of the $\frac{1}{q}$-Laughlin state
\cite{KA93}.
The states with
higher angular momenta are attributed to the edge of the
system \cite{Mac90a}.\\
The first proposal for a quasi-electron wave functions
is due to Laughlin \cite{Lau83b}:
\begin{equation}
\Psi ^{(+,z_{0})}_{\frac{1}{q}}=\prod_{i=1}^{N} e^{-
\frac{1}{4}|z_{i}|^{2}}(2\partial_{z_{i}}-z_{0}^{*})\prod_{k<l}^{N}(z_{k}-
z_{l})^{q}\:.
\end{equation}
This trial wave function can either be viewed as a quasi-electron located at
a position $z_{0}$ or it can be expanded with respect to $z_{0}$ in order to
get components, which are eigenfunctions of the total angular momentum
varying from $M^{*}-N$ to $M^{*}$ (the state with $M^{*}-1$ is missing).
All components of the
Laughlin quasi-electron wave function, except the one at angular momentum
$M^{*}-N$, contain Slater determinants with single particle angular momentum
$3(N-1)$ and thus violate the constraint (2). Only in the component at $M^{*}-
N$ all derivatives act on the polynomial part in (3) and, thus, here
the constraint (2) is met. Therefore,
we have also studied a modified Laughlin quasi-electron trial wave function
$\tilde{\Psi}^{(+,z_{0})}_{\frac{1}{q}}$, in which all Slater
determinants containing single
particle angular momenta $m>m_{max}$ are excluded. This wave
function has the
correct filling factor, but one gets higher energy. \\
The next quasi-electron trial wave function was proposed by MacDonald and
Girvin \cite{MG86a}. It is written in occupation number representation:
\begin{equation}
|\phi_{\frac{1}{q}}^{(+)}> = \hat{d}_{\infty} \ldots \hat{d_{1}}(1-
\hat{n}_{0})|\Psi _{\frac{1}{q}}>.
\end{equation}
The operator defined as $\hat{d}_{m}=c_{m-1}^{+}c_{m} +1 -n_{m} $
decreases
all one-particle angular momenta
by one. All states containing zero one-particle
angular momentum are projected out by $(1-\hat{n}_{0})$.
Unfortunately, one obtains only a single component (with
$M=M^{*}-N$).\\
As already mentioned, the third trial function is due to Jain \cite{Jai89b}.
Its component with angular momentum $M^{*}-N+m+1$ ($m=-1,0,\ldots,N-2$) is
given by
\begin{equation}
\chi_{\frac{1}{3}}^{(+,M^{*}-N+m+1)}={\cal P}_{0}\prod_{i<j}^{N}
(z_{i}-z_{j})^{2}
\left| \begin{array}{ccc}
z_{1}^{m}|z_{1}|^{2} & \ldots &
z_{N}^{m}|z_{N}|^{2} \\
1 & \ldots & 1 \\
z_{1}  & \ldots & z_{N} \\
\vdots & \ddots & \vdots \\
z_{1}^{N-2} & \ldots & z_{N}^{N-2}
\end{array} \right|
\prod_{i=1}^{N}e^{-\frac{1}{4}|z_{i}|^{2}}  \; .
\end{equation}
${\cal P}_{0}$ is the above mentioned projector onto the lowest Landau
level.\\
A fourth proposal based on the projection of the trial wave function from the
sphere onto the plane was made by us \cite{KA92}.
This method generates $ N+1$ angular momentum components resulting from the
expansion of
\begin{equation}
\Phi _{\frac{1}{q}}^{(+,z_{0})}
=\prod_{i=1}^{N} ((N-1)q-z_{i}\partial_{z_{i}}+z_{0}
\partial_{z_{i}}) \Psi _{\frac{1}{q}}\:
\end{equation}
with respect to $z_{0}$. In (6), the derivatives act as usual
only on the polynomial part of $\Psi_{\frac{1}{q}}$.
All the components of (6) have the correct filling factor (condition (2) is
fulfilled).\\
Expanding the components of all trial wave functions (3)-(6) in Slater
determinants, we calculated numerically exact the coefficients
for particle numbers up to $N=10$. From these coefficients, then
the energy expectation values were determined for $\Psi^{(+)}_{\frac{1}{3}},
\tilde{\Psi }^{(+)}_{\frac{1}{3}}, \phi^{(+)}_{\frac{1}{3}},
\chi^{(+)}_{\frac{1}{3}}$ and
$\Phi^{(+)}_{\frac{1}{3}}$. The results are shown for $N=8, 9, 10$ in
Table 1a, b and c, respectively. The data are pictured in Fig.\ 1 for $N$=8. \\
All data for $N=8,9,10$ show a very similar behaviour with respect to
their absolute values as well as to their dependence on the angular momentum
$M$.
In the case of 8
particles we added, for the sake of comparison, the lowest energy for each
$M$ resulting from the exact
numerical diagonalization of $H$, where the maximum single particle angular
momentum is taken to be $m_{max}=3(N-1)-1$.
It is quite obvious that the exact energies do not depend strongly on $M$
and this supports the picture of degeneracy for $M=M^{*}-
N$ to $M^{*}-1$. The absolute values of the data are near the
quasi-electron energy at
$\nu =\frac{1}{3}$: 0.1905, which was estimated by extrapolation of
the finite-size
data from calculations on the sphere, see Table 2 in \cite{GM90}. The
differences between the energies of all
the trial wave functions and the exact energies
increase with increasing $M$.  \\
For all $N$ and all $M$-components, Jain's
quasi-electron wave function is most favorable. The $M$-dependences of our
and Jain's proposal parallel each other, but our data are always
larger. The difference between the energies of Laughlin's proposal (modified
to meet (2), second column in Table 1)
and the ones of Jain's
proposal increases with increasing $M$ for all $N$. We include in the Tables
also
the energy expectation values of Laughlin's original proposal (column 1).
These energy expectation values differ only about 2\% from the
ones of the modified Laughlin trial wave function which meets the condition
(2). Therefore, while the objection that Laughlin's original quasi-electron
does not show the correct filling factor ($m_{max}$) is certainly true, its
energies are not so much affected by the violation of the condition (2). \\
In the last row of the Tables,
we include the
energies for the wave functions with angular momentum $M^{*}$. These
data are well separated from those of the other angular momenta,
e.\ g.\ the expectation value for $N=10$ and $M=135$ in Table 1c is much lower
than all the other ones. Thus, we identify only the $N$ lowest $M$-components
with the quasi-electron state, cf.\  \cite{KA93}. \\
For $N=8$, we calculate the mean of the $N$ quasi-electron components from
 Laughlin's ($N-1$ values), Jain's and our trial
wave functions and from the exact data. We
find  0.3554, 0.3040, 0.3240 and 0.1942, respectively. Thus, the relative
deviations of the data of the three trial wave functions from the exact ones
are 83\%, 57\% and 66\%. Jain's proposal is closest to the exact value, but
it is not as good as the corresponding proposal in the spherical geometry. In
general, all the energies of the trial wave functions are still rather high so
that there is still room for improvement.\\
In conclusion, the quasi-electron on the disk can be identified in the
energy spectrum. There are
$N$ angular momentum components. The various proposals for trial
wave functions show rather different
behaviour and are of quite different quality. Our approach of studying all
the angular momentum components of the quasi-electron in the disk geometry
leads to a very detailed picture of the behaviour of the various proposals
for trial wave functions and allows for a clear distinction of the relative
merits of these.
\vspace{0.3cm}
\\We are grateful for stimulating discussions with J.\ K.\ Jain, A.\ H.\
MacDonald and W.\ Weller. We acknowledge partial support by the Deutsche
Forschungsgemeinschaft via grant Ap47/1-1 and by the European Community
within the SCIENCE-project, grant SCC$^{*}$-CT90-0020.
\newpage
\underline{Table 1a}   \vspace{0.2cm} \\
\begin{tabular}[t]{|c|c|c|c|c|c||c|}
\hline
M & $\Psi_{\frac{1}{3}}^{(+)}$ & $\tilde{\Psi}_{\frac{1}{3}}^{(+)}$
 & $\phi_{\frac{1}{3}}^{(+)}$ &
 $\chi _{\frac{1}{3}}^{(+)}$ & $\Phi_{\frac{1}{3}}^{(+)}$
& num. diagon. \\   \hline
76 & 0.2413 & 0.2413 & 0.2362 & 0.2033 & 0.2413 & 0.1903 \\  \hline
77 & 0.2191 & 0.2199 &   --   & 0.1932 & 0.2200 & 0.1828 \\  \hline
78 & 0.2467 & 0.2491 &   --   & 0.2246 & 0.2474 & 0.1982 \\  \hline
79 & 0.3141 & 0.3183 &   --   & 0.2799 & 0.2980 & 0.1964 \\  \hline
80 & 0.3966 & 0.4022 &   --   & 0.3319 & 0.3445 & 0.1968 \\  \hline
81 & 0.4879 & 0.4954 &   --   & 0.3737 & 0.3835 & 0.1955 \\  \hline
82 & 0.5822 & 0.5920 &   --   & 0.4034 & 0.4159 & 0.1954 \\  \hline
83 &   --   &  --    &   --   & 0.4216 & 0.4411 & 0.1983 \\  \hline\hline
84 & 0      & 0.0137 &   --   & 0.0169 & 0.0896 & 0.0132 \\  \hline
\end{tabular} \\                        \vspace{0.5cm} \\
\underline{Table 1b}                    \vspace{0.2cm} \\
\begin{tabular}[t]{|c|c|c|c|c|c|}
\hline
M & $\Psi_{\frac{1}{3}}^{(+)}$ & $\tilde{\Psi}_{\frac{1}{3}}^{(+)}$
 & $\phi_{\frac{1}{3}}^{(+)}$ &
 $\chi _{\frac{1}{3}}^{(+)}$ & $\Phi_{\frac{1}{3}}^{(+)}$     \\ \hline
99  & 0.2467 & 0.2467 & 0.2438  & 0.2065 & 0.2467 \\  \hline
100 & 0.2232 & 0.2239 &   --    & 0.1926 & 0.2240 \\  \hline
101 & 0.2284 & 0.2302 &   --    & 0.2047 & 0.2314 \\  \hline
102 & 0.2715 & 0.2747 &   --    & 0.2447 & 0.2688 \\  \hline
103 & 0.3379 & 0.3423 &   --    & 0.2943 & 0.3141 \\  \hline
104 & 0.4142 & 0.4198 &   --    & 0.3392 & 0.3546 \\  \hline
105 & 0.5000 & 0.5072 &   --    & 0.3762 & 0.3897 \\  \hline
106 & 0.5857 & 0.5947 &   --    & 0.4028 & 0.4192 \\  \hline
107 &   --   &  --    &   --    & 0.4202 & 0.4437 \\  \hline  \hline
108 & 0      & 0.0119 &   --    & 0.0148 & 0.0926 \\  \hline
\end{tabular}          \vspace{0.5cm} \\
\underline{Table 1c} \vspace{0.2cm} \\
\begin{tabular}[t]{|c|c|c|c|c|c|}
\hline
M & $\Psi_{\frac{1}{3}}^{(+)}$ & $\tilde{\Psi}_{\frac{1}{3}}^{(+)}$
 & $\phi_{\frac{1}{3}}^{(+)}$ &
 $\chi _{\frac{1}{3}}^{(+)}$ & $\Phi_{\frac{1}{3}}^{(+)}$     \\ \hline
125 & 0.2495 & 0.2495 & 0.2463 & 0.2082 & 0.2495 \\ \hline
126 & 0.2309 & 0.2316 & --     & 0.1960 & 0.2316 \\ \hline
127 & 0.2221 & 0.2235 & --     & 0.1954 & 0.2255 \\ \hline
128 & 0.2440 & 0.2464 & --     & 0.2191 & 0.2471 \\ \hline
129 & 0.2931 & 0.2966 & --     & 0.2603 & 0.2861 \\ \hline
130 & 0.3566 & 0.3610 & --     & 0.3047 & 0.3266 \\ \hline
131 & 0.4290 & 0.4345 & --     & 0.3449 & 0.3633 \\ \hline
132 & 0.5098 & 0.5164 & --     & 0.3780 & 0.3953 \\ \hline
133 & 0.5890 & 0.5972 & --     & 0.4024 & 0.4227 \\ \hline
134 & --     & --     & --     & 0.4189 & 0.4462 \\ \hline  \hline
135 & 0      & 0.0105 & --     & 0.0132 & 0.0956 \\ \hline
\end{tabular}
\newpage
\underline{Captions:}
\vspace{0.5cm} \\
Table 1(a-c): The energy expectation values for the trial wave functions
(3)-(6) for $N=8$ (1a), $N=9$ (1b) and $N=10$ (1c) particles in dependence on
angular momentum $M$. For $N=8$, also the results of the exact
diagonalization are given.
\vspace{0.5cm}  \\
Fig.\ 1: The energy expectation values of Laughlin's (dashed line), Jain's
(solid line) and our (broken line) quasi-electron trial wave function
plotted against
angular momentum $M$ for $N=8$ particles. The dotted lines correspond to the
exact values. The data are given in Table 1a.
\newpage
\bibliographystyle{physrev}

\end{document}